%% file: prl_VQ.tex
\documentclass[aps,prl,twocolumn,showpacs,superscriptaddress,groupedaddress]{revtex4}  
\usepackage{graphicx}  
\usepackage{dcolumn}   
\usepackage{bm}        
\usepackage{amssymb}   

\def\MET{{\mbox{$E\kern-0.57em\raise0.19ex\hbox{/}_{T}~$}}}

\begin{document}

\hspace{5.2in} \mbox{FERMILAB-PUB-10/409-E}

\title{Search for single vector-like quarks in $\mathbf{p\bar{p}}$ collisions at $\mathbf{\sqrt{s}=1.96}$~TeV}
\input author_list.tex       
\date{December 8, 2010}

\begin{abstract}
We present a search for hypothetical vector-like quarks in $p\bar{p}$ collisions at 
$\sqrt{s}=1.96$ TeV. The data were collected
by the D0 detector at the Fermilab Tevatron Collider and correspond to an integrated luminosity of 5.4~fb$^{-1}$. 
We select events with a final state composed of a $W$ or $Z$ boson and a jet consistent with a heavy object decay.
We observe no
significant excess in comparison to the background prediction
and set limits on production cross sections for vector-like quarks decaying to
$W$+jet and $Z$+jet. 
These are the most stringent limits to date for electroweak single vector-like quark
production at hadron colliders.
\end{abstract}

\maketitle


The standard model (SM) of particle physics, despite its many successes
in accurately describing interactions below the TeV scale, is known to
suffer shortcomings at higher energy scales. A wide range of theories
have therefore been proposed to describe phenomena at the TeV scale and beyond,
among them warped extra dimensions~\cite{WED_seed}, universal extra dimensions~\cite{UED_seed},
and little Higgs~\cite{LH_seed} models. There exist particular realizations of each of these
theories that predict the existence of vector-like quarks 
\cite{vecquarks_model,ued_fourthgen,little_higgs}, massive particles which share 
many characteristics with SM quarks. The notable exception is that 
the right-handed and left-handed components of vector-like quarks transform
in the same way under $SU(3)\times SU(2)\times U(1)$, hence their name.

\begin{figure}
\includegraphics[scale=0.15]{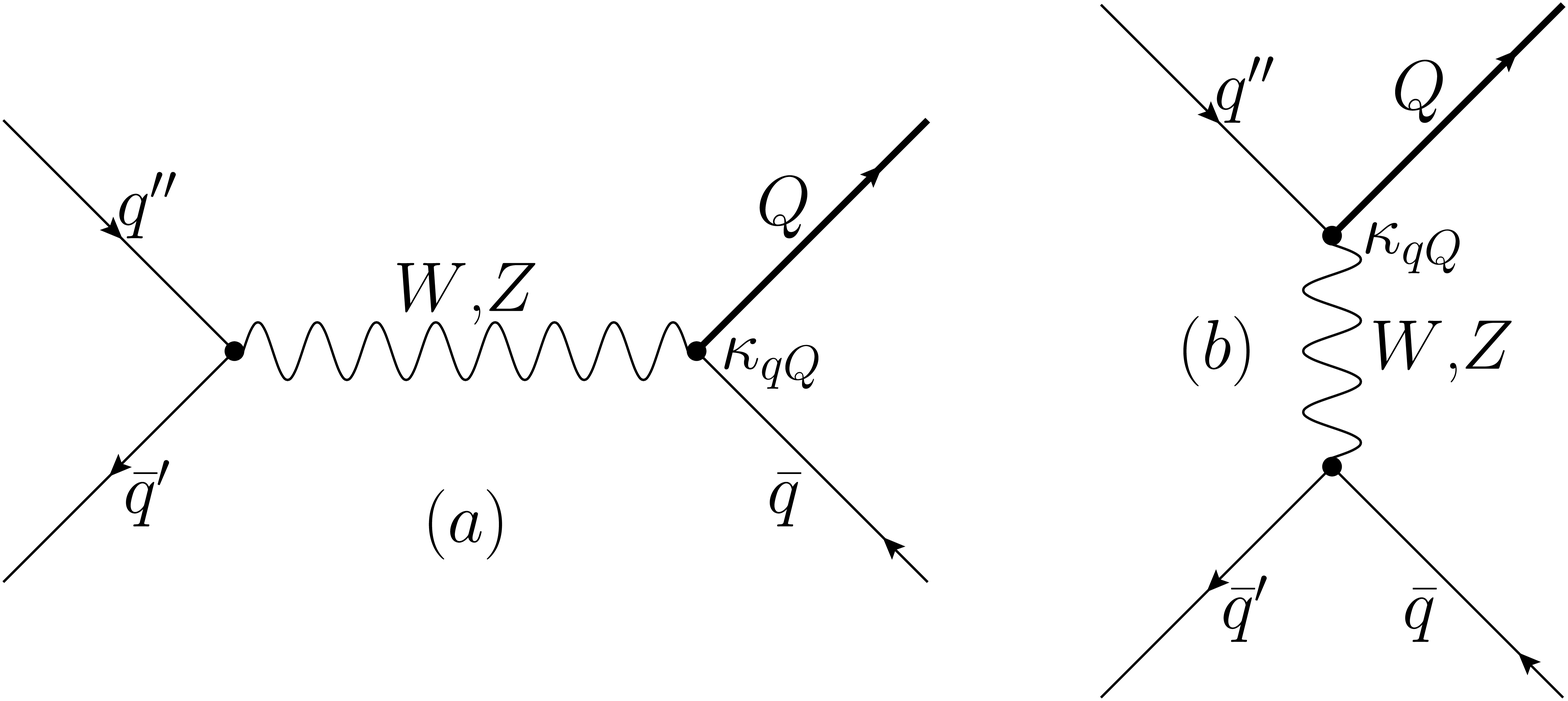}
\caption{\label{vq_prod}$s$-channel (a) and $t$-channel (b) electroweak production of a 
vector-like quark $Q$ at the Tevatron.}
\end{figure}

Previous searches for strong pair production of vector-like quarks at 
the Fermilab Tevatron Collider have excluded masses of up to 338 GeV
at the 95\% C.L.~\cite{cdf_tprime}. 
However, recent years have seen
the development of theoretical scenarios in which corrections to 
SM quark couplings arising from their
mixing with vector-like quarks cancel out, permitting the two types of quarks to mix 
to a degree unconstrained by precision electroweak measurements and $b$-factory
results. Such
large mixing with no introduction of anomalous couplings allows for single production
of vector-like quarks at hadron colliders via the weak interaction~\cite{vecquarks_model}.
Diagrams for single electroweak vector-like quark production are shown in Figure~\ref{vq_prod}.

Generally, electroweak couplings to SM quarks are set by the parameter
$\kappa_{qQ}$:

\begin{equation}
\label{vq_coupling}
\kappa_{qQ} = \frac{v}{m_{Q}}\tilde{\kappa}_{qQ},
\end{equation}

\noindent
where $v$ is the vacuum expectation value of the SM Higgs field, $m_{Q}$ is the mass 
of the vector-like quark, and $\tilde{\kappa}_{qQ}$ is the coupling strength.
Provided this
coupling is not too small, a search for singly produced vector-like quarks can benefit
from the lower kinematic threshold compared to pair production and stands to improve 
the mass limit considerably. 

In this Letter, we present a search for singly produced vector-like quarks
using data corresponding to an integrated luminosity of 5.4 fb$^{-1}$ collected 
by the D0 detector~\cite{run2det} at 
the Fermilab Tevatron Collider. 
We consider a model~\cite{vecquarks_model} in which 
there are two doublets of vector-like quarks that couple to the
first generation of SM quarks. Vector-like quark final states are characterized by either 
a $W$~or~$Z$ boson and at least two jets, one of which results from the
decay of the vector-like quark, while another is produced in association with
the vector-like quark at the primary vertex. We select events in which the vector 
boson decays to either electron(s) or muon(s).

The D0 detector is described in detail in~\cite{run2det}. The region of the D0 detector closest to the interaction 
point contains a central-tracking system consisting of a
silicon microstrip tracker (SMT) and a central fiber tracker (CFT),
both of which are located within a 2~T superconducting solenoidal
magnet. Hits in these two detectors are used to reconstruct tracks from charged
particles. Surrounding the two tracking subdetectors are
liquid-argon and uranium calorimeters, both electromagnetic and hadronic, which have a
central section (CC) covering pseudorapidities~\cite{d0coords}~$|\eta|\lesssim 1.1$, and two end calorimeters (EC) that extend coverage
to $|\eta|\approx 4.2$, with all three housed in separate
cryostats. Electrons are identified as isolated
energy clusters in the electromagnetic calorimeter, with a shape consistent
with that of an electron, matched to a track in the inner detector. Jets are reconstructed in the 
calorimeters using the iterative midpoint cone algorithm~\cite{midpoint} with a cone of radius ${\cal R}=0.5$ in $\eta-\phi$ space. An outer muon system, providing coverage for $|\eta|<2$,
consists of a layer of tracking detectors and scintillation trigger
counters in front of 1.8~T toroids, followed by two similar layers
after the toroids. Muons are identified via a combination of 
reconstructed tracks in both the outer muon system and the central tracking
system, and are required to be isolated from both tracker and calorimeter
activity.

Vector-like quark signals and background processes which include 
electrons or muons from 
$W$~or~$Z$~boson decays are modeled using Monte Carlo (MC) simulation. 
Signal samples are generated using {\sc madgraph}~\cite{madgraph}, with
CTEQ6L1~\cite{cteq} parton distribution functions, LO cross sections from~\cite{vecquarks_model}
and vector-like 
quark resonance widths calculated with {\sc bridge}~\cite{bridge}. 
Subsequent parton shower evolution is generated with 
{\sc pythia}~\cite{pythia}. 
We set $\tilde{\kappa}_{uD}=1, \tilde{\kappa}_{uU}=\sqrt{2}$ and 
$\tilde{\kappa}_{dU}=\tilde{\kappa}_{dD}=0$, such that down-type
vector-like quarks decay exclusively to $Wq$ and up-type quarks decay
exclusively to $Zq$. This choice, while affecting the production rates
and therefore the excluded vector-like quark masses, does not affect the 
cross section limits themselves.
We generate  up- and down-type vector-like quark signals with masses between 280 GeV and 700 GeV.
Backgrounds from $t\bar{t}$, $W\rightarrow \ell\nu$+jets, and 
$Z\rightarrow \ell\ell$+jets production are modeled using {\sc alpgen}~\cite{alpgen}, also
interfaced to {\sc pythia}. For these events, we use a matching procedure to avoid 
double counting partons produced by {\sc alpgen} and those added by the showering
in {\sc pythia}~\cite{MLM}.
Diboson samples are generated with 
{\sc pythia}, and single top quark production is modeled using the 
{\sc comphep}~\cite{comphep} generator. All MC samples are passed through a 
{\sc geant}~\cite{geant} simulation of the D0 detector and are
overlaid with data events from randomly chosen beam crossings to simulate the effect of 
multiple $p\bar{p}$ interactions and detector noise. MC events are then reconstructed using the same software that is used for data reconstruction. The $W$+jets and $Z$+jets samples are 
normalized to the leading-log cross section reported by {\sc alpgen} 
times a $k$-factor calculated by {\sc MCFM}~\cite{MCFM}. The $t\bar{t}$ samples
are normalized to a next-to-next-to-leading order (NNLO) cross section
for $m_{t}=172.5$ GeV~\cite{tt_xsec}, and the diboson samples ($WW$, $WZ$ and $ZZ$) to the NLO
cross section predicted by {\sc MCFM}~\cite{MCFM2}. All simulated events are corrected for 
differences in trigger and reconstruction efficiencies between data and 
simulation.

We conduct the search in two channels, corresponding to vector-like quark
decays to $(W\rightarrow\ell\nu)$+jet and to $(Z\rightarrow\ell\ell)$+jet. In the $W$+jets channel, we initially select events
which have passed a single lepton trigger and 
contain exactly one electron ($|\eta|<1.1$) or muon 
($|\eta|<2.0$) with transverse momentum $p_{T}>20$~GeV, at least 
two jets with $p_{T}>20$~GeV and $|\eta|<2.5$, and missing transverse energy $E\kern-0.57em\raise0.19ex\hbox{/}_{T}$, corrected for the momentum of any muons in the event, greater than 15 GeV.
We also require
$2M^{W}_{T}+E\kern-0.57em\raise0.19ex\hbox{/}_{T} > 80$ GeV, where $M^{W}_{T}$ is the transverse 
mass of the lepton-$E\kern-0.57em\raise0.19ex\hbox{/}_{T}$ system, in order to suppress
the multijet background. In the $Z$+jets channel, we initially select events with 
exactly two electrons or muons in addition to at least two jets, all with
same $p_{T}$ thresholds as used in the $W$+jets channel. Events are selected by a mixture of single lepton, dilepton and lepton plus jets triggers.
Due to the low rate of background without real
$Z$ bosons, electrons are also accepted in the endcap electromagnetic calorimeters
$(1.5<|\eta|<2.5)$. The two leptons are required to have an invariant
mass between 70 and 110~GeV, i.e.\ consistent with that of a $Z$ boson. Additionally, we
require $E\kern-0.57em\raise0.19ex\hbox{/}_{T}<50$ GeV, as this channel contains only 
instrumental sources of missing transverse energy. 

The largest physics background to the single lepton channel is
$W(\rightarrow \ell\nu)$+jets production, with smaller contributions
from $Z$+jets, $t\bar{t}$, diboson, and single top quark processes. 
The main instrumental background arises from multijet events in which one of 
the jets is misidentified as 
a high-$p_{T}$ isolated lepton in the detector. We model this background using data events
which fail the calorimeter shower shape requirements for the electron or the 
isolation requirements for the muon, but pass all other selection
criteria. In the single electron channel, we estimate the relative
fraction of real electrons from $W$ boson decays and misidentified electrons from
jets by determining the efficiencies for each type
of event to pass a tighter selection. These efficiencies are calculated
in $Z\rightarrow ee$ data for real electrons and in a $E\kern-0.57em\raise0.19ex\hbox{/}_{T}<20$
GeV sample for misidentified electrons. In the single muon channel, 
we scale the events failing the muon isolation requirement to match the number 
of events obtained after subtracting the expected number of real $W$ bosons satisfying
the event selection from the number of events observed in data.

In the dilepton channel, $(Z\rightarrow\ell\ell)$+jets events dominate the SM background.
In order to correct the Monte Carlo for the small trigger inefficiency
in the data, we apply a global normalization determined
using a pure $Z\rightarrow\ell\ell$ sample. The ratio between the data yield
in this sample and the predicted yield from the inclusive $Z$ boson
cross section multiplied by the branching ratio is associated with the overall trigger
efficiency, and the Monte Carlo distributions are scaled by this ratio. 
The multijet
background is modeled using data events in which both leptons fail quality
criteria (for electrons) or isolation criteria (for muons), and is normalized
to the difference between data and MC in the dilepton mass window
between 40 and 70 GeV.

The signal contains events with a $W$ or $Z$ boson and jets, all with
high transverse momentum, as the decay products of a 
high mass resonance. 
We apply several kinematic selection criteria to select events of this type and
to minimize the contributions from SM background processes.
In the single lepton channel, we require lepton $p_{T}>50$~GeV, 
highest jet $p_{T}>100$~GeV, $E\kern-0.57em\raise0.19ex\hbox{/}_{T}>40(50)$~GeV for the 
muon (electron) channel, and that the separation in
azimuthal angle $\phi$ between the lepton and the $E\kern-0.57em\raise0.19ex\hbox{/}_{T}$ 
be less than 2. 
As the signal contains real $W$ bosons, we additionally require 
$M^{W}_{T}<150$~GeV. Finally, we exploit the relationship between
the lepton charge and the $\eta$ of the jet with the second-highest
$p_{T}$  in the signal
topology. This jet, which we assign to the SM 
quark produced in association with the vector-like
quark, is emitted, in general, into one of the forward regions of
the detector, and its direction is strongly correlated with the 
charge of the produced vector-like quark, and thus also with 
the charge of the lepton from its decay.
We therefore require
$Q_{\ell}\times\eta_{j_{2}}>0$, where $Q_{\ell}$ is the lepton charge
and $\eta_{j_{2}}$ is the $\eta$ of the jet with the second-highest $p_{T}$
in the event. This selection is efficient for 
the signal ($\approx85\%$), while reducing the SM background 
by roughly a factor of two. 

In the dilepton channel, we similarly select events with properties
characteristic of a heavy resonance decay to a $Z$ boson and a jet. 
We require the $p_{T}$ of the dilepton system to be greater than 
100 GeV, the spatial separation of the two leptons in $\eta-\phi$ space
($\Delta {\cal R}$) to be less than 2.0, and the leading jet $p_{T}$ to 
be greater than 100 GeV. 

\begin{figure}
\includegraphics[scale=0.42]{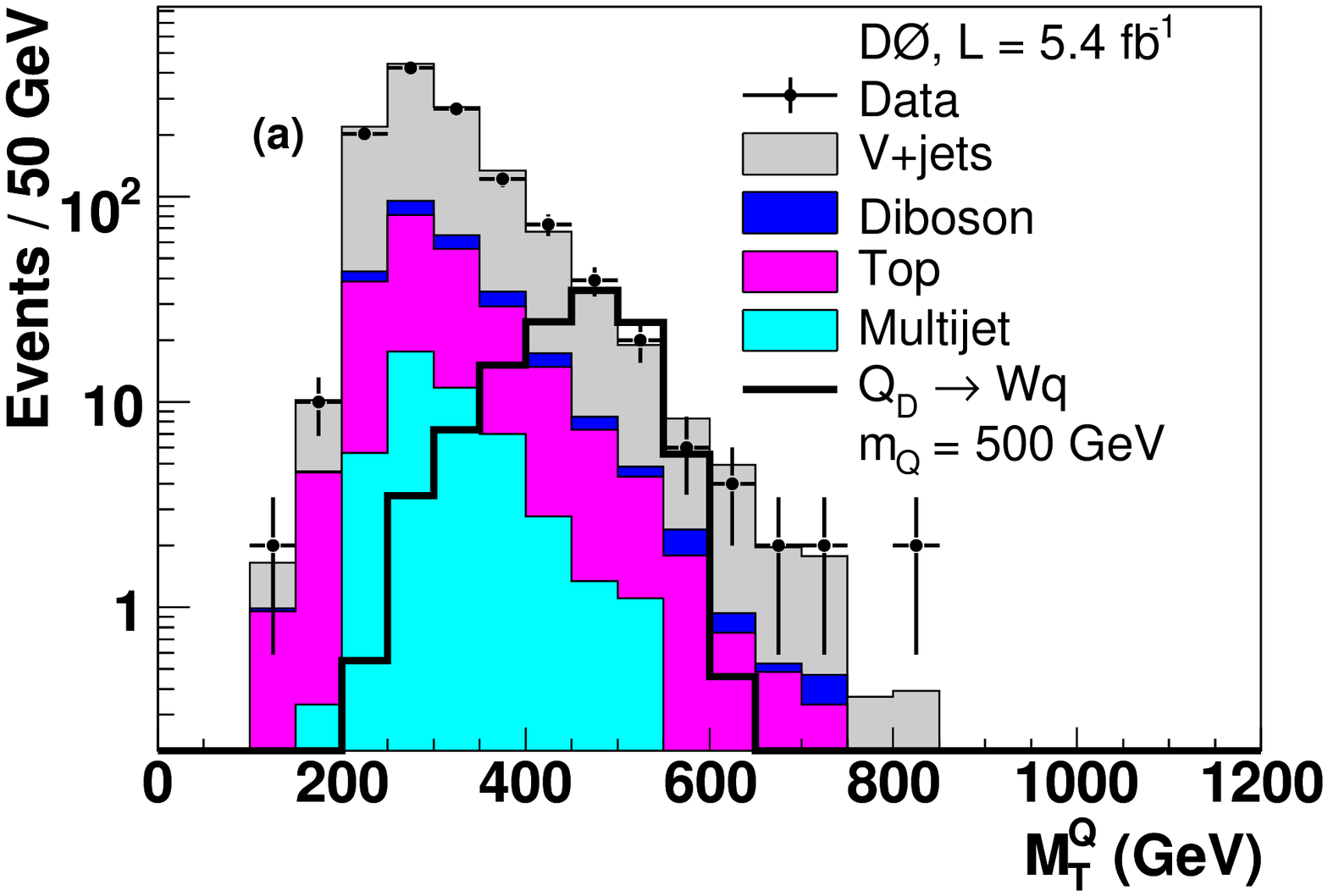}
\includegraphics[scale=0.42]{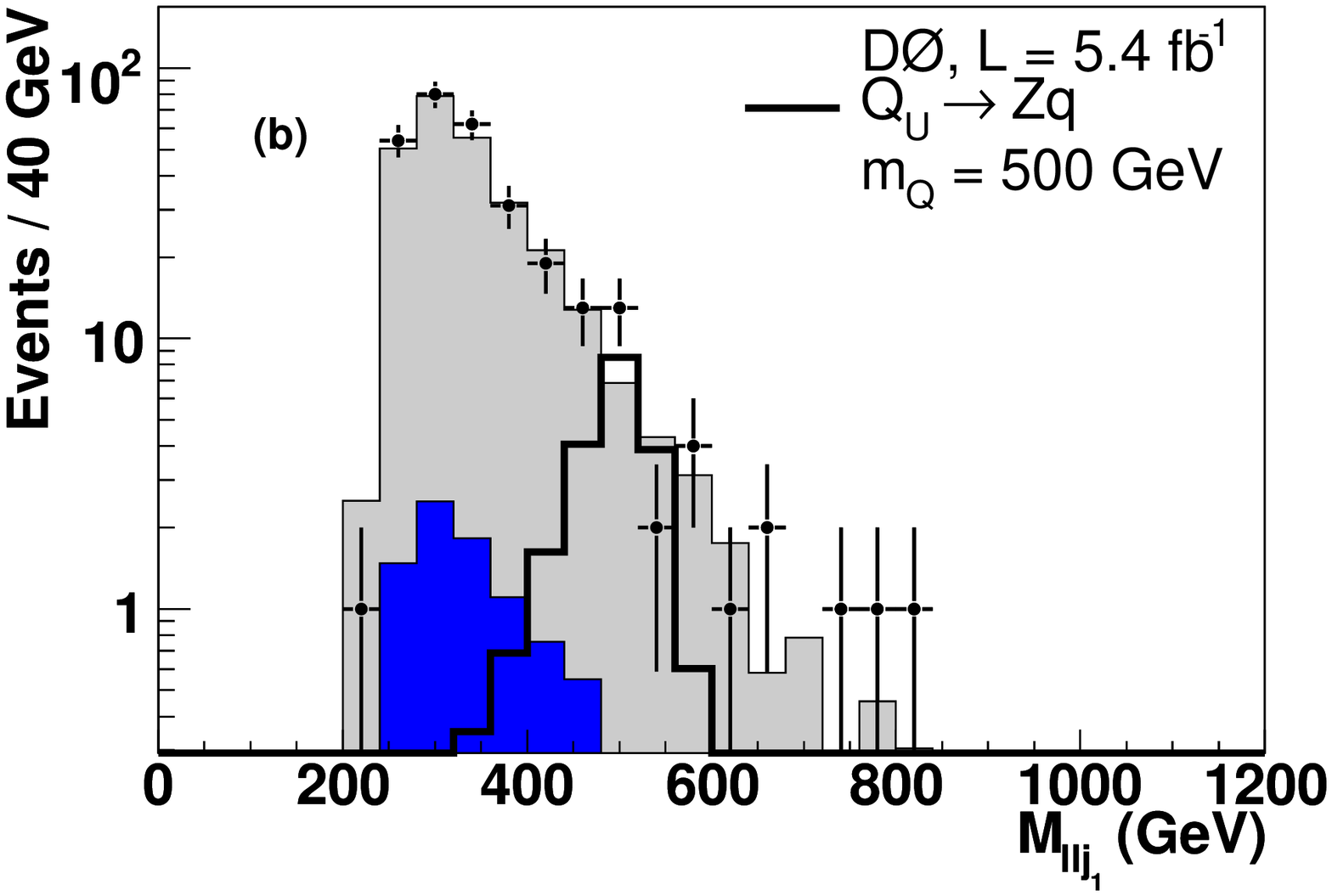}
\caption{\label{cluster_Mtr} (a) Vector-like quark transverse mass and 
(b) vector-like quark mass
for the single lepton and dilepton channels, respectively. Distributions for signal
processes $Q_{D}\rightarrow Wq$ and $Q_{U}\rightarrow Zq$ 
are normalized to the integrated luminosity of the data and include detector acceptance and
reconstruction efficiencies. These assume $\tilde{\kappa}_{uD}=1$ and
$\tilde{\kappa}_{uU}=\sqrt{2}$.}
\end{figure}

Table~\ref{final_yields} displays
the observed number of data events, 
along with estimated background yields, after the final
event selection in the single lepton and dilepton channels.
Fig.~\ref{cluster_Mtr} shows the reconstructed
vector-like quark transverse mass, defined as

\begin{equation}
\label{cluster_mtr_def}
(M_{T}^{Q})^{2}=(\sqrt{p_{T_{W}}^{2}+M_{W}^{2}}+p_{T_{j_{1}}})^{2}-(\vec{p}_{T_{W}}+\vec{p}_{T_{j_{1}}})^{2},
\end{equation}

\noindent
for the single lepton channel, where $M_{W}=80.3$~GeV is the mass of the $W$ boson and 
$\vec{p}_{T_{j1}}$ refers to the transverse
momentum of the leading jet in the event. Also shown in Fig.~\ref{cluster_Mtr}
is the vector-like quark mass in the dilepton channel, reconstructed as the invariant
mass of the dilepton + leading jet system.

\begin{table}
\caption{\label{final_yields}Predicted number of background events including
total uncertainties and observed number of data events after final selection.}
\begin{ruledtabular}
\begin{tabular}{lD{,}{\,\pm\,}{-1}D{,}{\,\pm\,}{-1}}
Process&\multicolumn{1}{c}{Single lepton sample}&\multicolumn{1}{r}{Dilepton sample}\\
\hline
Multijet    &    47.7, 4.7  &   < 0.1  \\
$Z$+jets        &    39.9, 7.4  &   262, 45  \\
$W$+jets        &    901, 159   &   0.3, 0.2  \\
Top             &    193, 24    &   0.57, 0.06\\
Diboson         &    38.6, 3.8  &   8.3, 0.7  \\
\hline
Background sum  &    1220, 161 &    271, 45\\
\hline
Data            &    1175  & 285 \\ 
\end{tabular}
\end{ruledtabular}
\end{table}

Major sources of systematic uncertainty common to both analyses 
include modeling of $W/Z$+jets backgrounds (15\%); cross
sections for $t\bar{t}$ (10\%), diboson (6\%), and $W$/$Z$+jets
(6\%) production; and jet energy scale and resolution (1\%--5\%).
Major systematic uncertainties unique to the
single lepton analysis include integrated luminosity (6.1\%)~\cite{d0lumi},
lepton identification efficiencies (3\%), high-$p_{T}$ muon modeling, and trigger modeling (1\%). 
A systematic uncertainty of 5\% is 
assigned to the global background normalization applied to the dilepton MC
samples. Systematic uncertainties
on multijet normalization are 6.5\%--100\%, depending on the
channel. These do not have a large effect on the overall 
background prediction, as the estimated multijet background is small after
the final selection.

\begin{figure}
\includegraphics[scale=0.42]{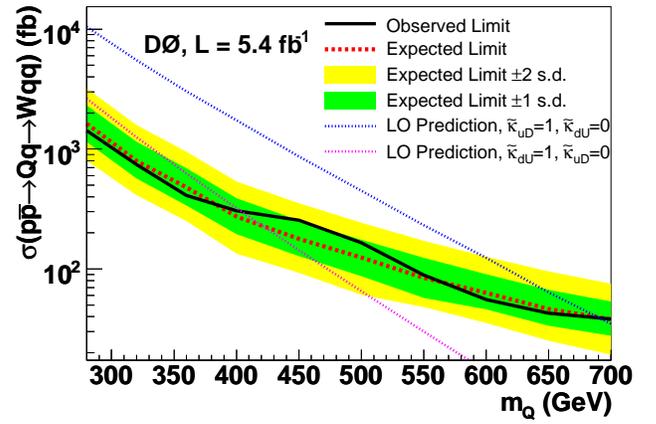}
\caption{\label{single_lepton_limit} Excluded production cross section
for a vector-like quark $Q$ decaying to $Wq$ as a function of $m_Q$, compared to LO predictions of vector-like quark production with different $\tilde{\kappa}_{qQ}$.}
\end{figure}

\begin{figure}
\includegraphics[scale=0.42]{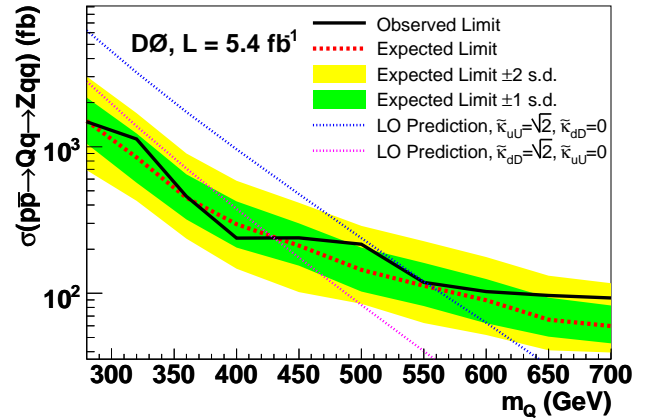}
\caption{\label{dilepton_limit} Excluded production cross section
for a vector-like quark $Q$ decaying to $Zq$ as a function of $m_Q$, compared to LO predictions of vector-like quark production with different $\tilde{\kappa}_{qQ}$.}
\end{figure}

We observe no significant excess of data over the background
prediction in either channel, therefore we set limits on 
vector-like quark production cross sections. 
We employ a modified frequentist approach using the 
D0 likelihood fitter~\cite{collie}
that incorporates a log-likelihood ratio $(LLR)$ test statistic~\cite{CLs}.
We calculate confidence levels for the signal plus background 
$(CL_{s+b})$ and background-only $(CL_{b})$ hypotheses by integrating
the $(LLR)$ distributions obtained by generating pseudo-experiments
using Poisson statistics. Systematic uncertainties are treated as 
uncertainties on the expected number of signal and background events.
The $95\%$ confidence level limit is defined as the 
signal cross section for which $CL_{s} \equiv CL_{s+b}/CL_{b}$ is $0.05$.
Using as discriminant
variables the vector-like quark transverse mass for decays to $Wq$ and 
the vector-like quark mass
for decays to $Zq$, we obtain 95\% C.L. limits on vector-like
quark production cross sections. These are shown in Figs.~\ref{single_lepton_limit}
and~\ref{dilepton_limit}, along with leading order theoretical predictions for 
two different scenarios.  
For the case of $\tilde{\kappa}_{uD}=1$ and $\tilde{\kappa}_{uU}=\sqrt{2}$ with no coupling
to the down quark, we exclude
masses below 693~GeV for a vector-like quark decaying exclusively to $Wq$ and
masses below 551~GeV for a vector-like quark decaying exclusively to $Zq$ at the 95\% C.L.
For an alternate scenario defined by $\tilde{\kappa}_{dU}=1$ and $\tilde{\kappa}_{dD}=\sqrt{2}$
with no coupling to the up quark, the corresponding mass limits are 403 GeV and 430 GeV, respectively.

In summary, we have presented a search for single vector-like quark production
at the Tevatron in the $W$+jets and $Z$+jets final states. The observed data
are consistent with the background expectation and limits on vector-like
quark cross sections are derived. We exclude vector-like quark masses below
693~GeV for decays to $Wq$ and masses below 551~GeV for decays to $Zq$
at $95\%$ C.L.
assuming vector-like quark -- SM quark mixing
parameters $\tilde{\kappa}_{uD}=1$ and $\tilde{\kappa}_{uU}=\sqrt{2}$ along with $100\%$ branching ratios to
the respective final states.

\input acknowledgement.tex   

We additionally thank Marcela Carena and Anupama Atre of the Fermi National
Accelerator Laboratory for providing the {\sc madgraph} implementation
of the vector-like quarks model.

\end{document}

%% file: author_list.tex
\affiliation{Universidad de Buenos Aires, Buenos Aires, Argentina}
\affiliation{LAFEX, Centro Brasileiro de Pesquisas F{\'\i}sicas, Rio de Janeiro, Brazil}
\affiliation{Universidade do Estado do Rio de Janeiro, Rio de Janeiro, Brazil}
\affiliation{Universidade Federal do ABC, Santo Andr\'e, Brazil}
\affiliation{Instituto de F\'{\i}sica Te\'orica, Universidade Estadual Paulista, S\~ao Paulo, Brazil}
\affiliation{Simon Fraser University, Vancouver, British Columbia, and York University, Toronto, Ontario, Canada}
\affiliation{University of Science and Technology of China, Hefei, People's Republic of China}
\affiliation{Universidad de los Andes, Bogot\'{a}, Colombia}
\affiliation{Charles University, Faculty of Mathematics and Physics, Center for Particle Physics, Prague, Czech Republic}
\affiliation{Czech Technical University in Prague, Prague, Czech Republic}
\affiliation{Center for Particle Physics, Institute of Physics, Academy of Sciences of the Czech Republic, Prague, Czech Republic}
\affiliation{Universidad San Francisco de Quito, Quito, Ecuador}
\affiliation{LPC, Universit\'e Blaise Pascal, CNRS/IN2P3, Clermont, France}
\affiliation{LPSC, Universit\'e Joseph Fourier Grenoble 1, CNRS/IN2P3, Institut National Polytechnique de Grenoble, Grenoble, France}
\affiliation{CPPM, Aix-Marseille Universit\'e, CNRS/IN2P3, Marseille, France}
\affiliation{LAL, Universit\'e Paris-Sud, CNRS/IN2P3, Orsay, France}
\affiliation{LPNHE, Universit\'es Paris VI and VII, CNRS/IN2P3, Paris, France}
\affiliation{CEA, Irfu, SPP, Saclay, France}
\affiliation{IPHC, Universit\'e de Strasbourg, CNRS/IN2P3, Strasbourg, France}
\affiliation{IPNL, Universit\'e Lyon 1, CNRS/IN2P3, Villeurbanne, France and Universit\'e de Lyon, Lyon, France}
\affiliation{III. Physikalisches Institut A, RWTH Aachen University, Aachen, Germany}
\affiliation{Physikalisches Institut, Universit{\"a}t Freiburg, Freiburg, Germany}
\affiliation{II. Physikalisches Institut, Georg-August-Universit{\"a}t G\"ottingen, G\"ottingen, Germany}
\affiliation{Institut f{\"u}r Physik, Universit{\"a}t Mainz, Mainz, Germany}
\affiliation{Ludwig-Maximilians-Universit{\"a}t M{\"u}nchen, M{\"u}nchen, Germany}
\affiliation{Fachbereich Physik, Bergische  Universit{\"a}t Wuppertal, Wuppertal, Germany}
\affiliation{Panjab University, Chandigarh, India}
\affiliation{Delhi University, Delhi, India}
\affiliation{Tata Institute of Fundamental Research, Mumbai, India}
\affiliation{University College Dublin, Dublin, Ireland}
\affiliation{Korea Detector Laboratory, Korea University, Seoul, Korea}
\affiliation{CINVESTAV, Mexico City, Mexico}
\affiliation{FOM-Institute NIKHEF and University of Amsterdam/NIKHEF, Amsterdam, The Netherlands}
\affiliation{Radboud University Nijmegen/NIKHEF, Nijmegen, The Netherlands}
\affiliation{Joint Institute for Nuclear Research, Dubna, Russia}
\affiliation{Institute for Theoretical and Experimental Physics, Moscow, Russia}
\affiliation{Moscow State University, Moscow, Russia}
\affiliation{Institute for High Energy Physics, Protvino, Russia}
\affiliation{Petersburg Nuclear Physics Institute, St. Petersburg, Russia}
\affiliation{Stockholm University, Stockholm and Uppsala University, Uppsala, Sweden }
\affiliation{Lancaster University, Lancaster LA1 4YB, United Kingdom}
\affiliation{Imperial College London, London SW7 2AZ, United Kingdom}
\affiliation{The University of Manchester, Manchester M13 9PL, United Kingdom}
\affiliation{University of Arizona, Tucson, Arizona 85721, USA}
\affiliation{University of California Riverside, Riverside, California 92521, USA}
\affiliation{Florida State University, Tallahassee, Florida 32306, USA}
\affiliation{Fermi National Accelerator Laboratory, Batavia, Illinois 60510, USA}
\affiliation{University of Illinois at Chicago, Chicago, Illinois 60607, USA}
\affiliation{Northern Illinois University, DeKalb, Illinois 60115, USA}
\affiliation{Northwestern University, Evanston, Illinois 60208, USA}
\affiliation{Indiana University, Bloomington, Indiana 47405, USA}
\affiliation{Purdue University Calumet, Hammond, Indiana 46323, USA}
\affiliation{University of Notre Dame, Notre Dame, Indiana 46556, USA}
\affiliation{Iowa State University, Ames, Iowa 50011, USA}
\affiliation{University of Kansas, Lawrence, Kansas 66045, USA}
\affiliation{Kansas State University, Manhattan, Kansas 66506, USA}
\affiliation{Louisiana Tech University, Ruston, Louisiana 71272, USA}
\affiliation{Boston University, Boston, Massachusetts 02215, USA}
\affiliation{Northeastern University, Boston, Massachusetts 02115, USA}
\affiliation{University of Michigan, Ann Arbor, Michigan 48109, USA}
\affiliation{Michigan State University, East Lansing, Michigan 48824, USA}
\affiliation{University of Mississippi, University, Mississippi 38677, USA}
\affiliation{University of Nebraska, Lincoln, Nebraska 68588, USA}
\affiliation{Rutgers University, Piscataway, New Jersey 08855, USA}
\affiliation{Princeton University, Princeton, New Jersey 08544, USA}
\affiliation{State University of New York, Buffalo, New York 14260, USA}
\affiliation{Columbia University, New York, New York 10027, USA}
\affiliation{University of Rochester, Rochester, New York 14627, USA}
\affiliation{State University of New York, Stony Brook, New York 11794, USA}
\affiliation{Brookhaven National Laboratory, Upton, New York 11973, USA}
\affiliation{Langston University, Langston, Oklahoma 73050, USA}
\affiliation{University of Oklahoma, Norman, Oklahoma 73019, USA}
\affiliation{Oklahoma State University, Stillwater, Oklahoma 74078, USA}
\affiliation{Brown University, Providence, Rhode Island 02912, USA}
\affiliation{University of Texas, Arlington, Texas 76019, USA}
\affiliation{Southern Methodist University, Dallas, Texas 75275, USA}
\affiliation{Rice University, Houston, Texas 77005, USA}
\affiliation{University of Virginia, Charlottesville, Virginia 22901, USA}
\affiliation{University of Washington, Seattle, Washington 98195, USA}
\author{V.M.~Abazov} \affiliation{Joint Institute for Nuclear Research, Dubna, Russia}
\author{B.~Abbott} \affiliation{University of Oklahoma, Norman, Oklahoma 73019, USA}
\author{B.S.~Acharya} \affiliation{Tata Institute of Fundamental Research, Mumbai, India}
\author{M.~Adams} \affiliation{University of Illinois at Chicago, Chicago, Illinois 60607, USA}
\author{T.~Adams} \affiliation{Florida State University, Tallahassee, Florida 32306, USA}
\author{G.D.~Alexeev} \affiliation{Joint Institute for Nuclear Research, Dubna, Russia}
\author{G.~Alkhazov} \affiliation{Petersburg Nuclear Physics Institute, St. Petersburg, Russia}
\author{A.~Alton$^{a}$} \affiliation{University of Michigan, Ann Arbor, Michigan 48109, USA}
\author{G.~Alverson} \affiliation{Northeastern University, Boston, Massachusetts 02115, USA}
\author{G.A.~Alves} \affiliation{LAFEX, Centro Brasileiro de Pesquisas F{\'\i}sicas, Rio de Janeiro, Brazil}
% ...
%
%
\collaboration{The D0 Collaboration\footnote{with visitors from
$^{a}$Augustana College, Sioux Falls, SD, USA,
$^{b}$The University of Liverpool, Liverpool, UK,
$^{c}$SLAC, Menlo Park, CA, USA,
$^{d}$ICREA/IFAE, Barcelona, Spain,
$^{e}$Centro de Investigacion en Computacion - IPN, Mexico City, Mexico,
$^{f}$ECFM, Universidad Autonoma de Sinaloa, Culiac\'an, Mexico,
and 
$^{g}$Universit{\"a}t Bern, Bern, Switzerland.%
}} \noaffiliation
\vskip 0.25cm

%% file: acknowledgement.tex
%
We thank the staffs at Fermilab and collaborating institutions,
and acknowledge support from the
DOE and NSF (USA);
CEA and CNRS/IN2P3 (France);
FASI, Rosatom and RFBR (Russia);
CNPq, FAPERJ, FAPESP and FUNDUNESP (Brazil);
DAE and DST (India);
Colciencias (Colombia);
CONACyT (Mexico);
KRF and KOSEF (Korea);
CONICET and UBACyT (Argentina);
FOM (The Netherlands);
STFC and the Royal Society (United Kingdom);
MSMT and GACR (Czech Republic);
CRC Program and NSERC (Canada);
BMBF and DFG (Germany);
SFI (Ireland);
The Swedish Research Council (Sweden);
and
CAS and CNSF (China).

%% file: prl_VQ.bbl
\begin{thebibliography}{99}
\bibitem{d0coords}
The D0 detector utilizes a right-handed coordinate system with the $z$~axis
pointing in the direction of the proton beam and the $y$~axis pointing upwards.
The azimuthal angle $\phi$~is defined in the $x$-$y$~plane measured from the $x$~axis.
The pseudorapidity is defined as $\eta=-\textup{ln}[\textup{tan}(\theta/2)]$, where
$\theta=\textup{arctan}(\sqrt{x^2+y^2}/z)$. Transverse variables are defined as
projections of the variables onto the $x$-$y$~plane.
\end{thebibliography}
